\documentclass[twocolumn,showpacs,preprintnumbers,superscriptaddress,
prb,amssymb]{revtex4}
\usepackage{graphicx}% Include figure files
\usepackage{dcolumn}% Align table columns on decimal point
\usepackage{bm}% bold math
\usepackage{amsmath}
\usepackage{epstopdf}
\usepackage{color}
\usepackage{ulem}

\newcommand{\eps}{\varepsilon}
\newcommand{\red}[1]{{\color{black} #1}}

\begin{document}

\title{Dephasing in single-electron generation due to environmental noise\\
probed by Hong Ou Mandel interferometry}
\author{Eiki Iyoda}
\email{iyoda@noneq.c.u-tokyo.ac.jp}
\affiliation{Department of Basic Science, University of Tokyo, 3-8-1 Komaba, Meguro, Tokyo 153-8902, Japan}
\author{Takeo Kato}
\affiliation{Institute for Solid State Physics, University of Tokyo, Kashiwa, Chiba 277-8581, Japan}
\author{Kazuki Koshino}
\affiliation{College of Liberal Arts and Sciences, 
Tokyo Medical and Dental University, Ichikawa, Chiba 272-0827, Japan}
\author{Thierry Martin}
\affiliation{\red{Aix Marseille Universit\'e, Universit\'e de Toulon, CNRS, CPT, UMR 7332, 13288 Marseille, France}}

\date{\today}

\begin{abstract} 
We consider the effect of dephasing on a quantum dot which injects single electrons on a chiral edge 
channel of the quantum Hall effect. 
Dephasing is described by the coupling of the dot to a bosonic bath which represents the electromagnetic 
environment. Using the input-output formalism of quantum optics, we derive the density matrix
of the edge degrees of freedom. Results are illustrated by computing the zero frequency current-current 
correlations when two such single electron emitters achieve a collision at the location of a quantum point 
contact, in the same spirit as the Hong Ou Mandel experiment of quantum optics. Such correlations are directly linked 
to the quantum mechanical purity. We show that as observed in a recent experiment, the effect of dephasing 
leads to a \red{lifting} of the Hong Ou Mandel dip when the time delay between the two electron wave 
packets is zero. Generalizations to time filtered wave packets as well as to asymmetric, detuned injection 
between opposite edges are obtained.      
\end{abstract}
\maketitle

\section{Introduction}

% 1. SEG, moving flying qubit, Bell pair,
Quantum mesoscopic physics, or nanophysics, aims at studying the manifestations of quantum mechanics, such as interference effects and coherence, with electron transport in condensed matter materials. Such manifestations have been studied in the context of quantum optics since the middle of the last century, where fundamental tests of quantum mechanics were explored, for instance in Hanbury Brown and Twiss\cite{HBT} (HBT) and Hong Ou Mandel\cite{Hong87} 
(HOM) experiments, more recently with single photon sources.\cite{single_photon} In nanophysics, there is a growing interest to translate these concepts of quantum optics to electrons propagating in nanostructures. To a large extent, this is due to the recent availability of on-demand single electron 
sources,\cite{Ahlers06,Blumenthal07,Feve07,Leicht11,Hermelin11,McNeil11} and of material which acts as wave guides for the electrons, such as edge states in the quantum Hall effect.  Electrons differ from photons because of their fermionic statistics, in condensed matter settings they are always accompanied by a Fermi sea, and as charged particles they interact strongly between themselves and with their environment. 

In electronic quantum optics, the fermionic counterpart of the Hong-Ou-Mandel(HOM) experiment\cite{Hong87}
has been considered theoretically in a setup of a two-electron collider,\cite{Olkhovskaya08,Moskalets11}
and has been realized quite recently in the integer quantum Hall effect regime\cite{Bocquillon13}
\red{and in a point contact system using levitons.\cite{Dubois13a}} 
Nonlocal quantum correlation between propagating electrons in conduction channels
and related nonlocal transport have also been studied theoretically 
towards applications to quantum information processing.\cite{Beenakker03,Splettstoesser09}

% 2. Evaluation, dephasing
The experiments of on-demand electron generation \red{in electronic transmission channels}
have stimulated the theoretical study on the quantum-mechanical nature of the single 
electrons which are generated \red{in the one-dimensional channel}.~\cite{Moskalets08,Splettstoesser08,Keeling08,Battista11,Haack11,Grenier11,
Jonckheere11,Jonckheere12,Moskalets13a,Moskalets13b,Haack13,Ferraro13,Grenier13,Dubois13b}
The excess current noise at the output of a quantum point contact (QPC), 
which includes the information about the coherence of generated electrons,
has been studied both theoretically and experimentally.\cite{Mahe10,Albert10,Parmentier12,Bocquillon12}
This coherence is directly measured by the degree of antibunching
in the fermonic HOM experiment, which reflects the indistinguishability of electrons.
The imperfect antibunching reported in Ref.~\onlinecite{Bocquillon13} (the fact that the HOM dip
\red{is lifted} for zero time delay between
electron wave packets) \red{includes information about the distinguishability of propagating 
electrons in the chiral edge channels before arriving at the collision point (the QPC)
as well as asymmetry of the wave packet due to difference in parameters
of two single-electron generator.}

At the present, the dephasing of propagating electrons within propagating channels
has been considered mostly by
phenomenological approaches,\cite{Degiovanni09} by ad-hoc fitting 
of the experimental curves,\cite{Bocquillon13} or alternatively using the bosonization 
approach.\cite{Feve08,Wahl13} In these works, the relaxation due to Coulomb interaction
between an edge state at its electromagnetic environment or between propagating edge 
channels\cite{Roulleau08a,Roulleau08b,Altimiras10a,Altimiras10b,Huynh12} 
is assumed to be the main source of decoherence.

% 3. in this paper,,,,
Yet, there are many other possibilities for other sources of decoherence, among them the simple fact that the 
metallic gates surrounding the dot represent a fluctuating electromagnetic environment.  
In this paper, we examine the effect of electron decoherence by focusing on the role of the energy-level fluctuations
of the quantum dot due to this environment. We present a simple framework for evaluating the quality of the
generated electrons reflecting indistinguishability of electrons based on the so-called input-output relation,
which is a standard tool in quantum optics.\cite{Yurke84,Collett84,Gardiner88,Walls95,Gardiner04,Iyoda13}
To our knowledge, the input-output relations have so far not been applied to a fermonic quantum optics setup.
Our calculation provides a useful and comprehensive picture of dephasing effects on the
quality of generated single electrons as well as simple formulae for the degree of
antibunching in the HOM experiment. Finally, we also show that the decoherence due to the energy-level 
fluctuation at the quantum dot can be reduced by a filtering technique.

\section{Model}

\begin{figure}[tb]
\begin{center}
\includegraphics[width=0.9 \columnwidth]{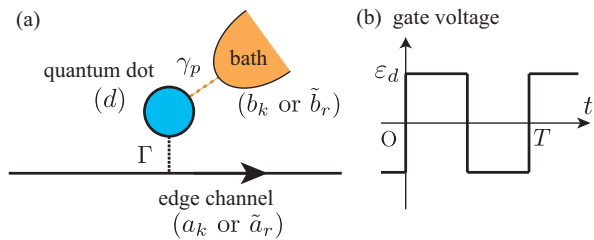}
\caption{(a) Schematic picture of the model considered in this paper: a quantum dot (blue) which is coupled
to an electromagnetic environement (orange) is subject to a periodic drive in order to transfer electrons
on the neighboring chiral edge state.  
(b) The square periodic voltage pulse which is applied to the quantum dot.}
\label{fig:model}
\end{center}
\end{figure}

% Model
In actual experiments, the single electron source consists of a mesoscopic capacitor\cite{gabelli} which is 
a rather ``large''\cite{footnote4} quantum dot, connected to a quantum Hall edge channel propagating in one direction. 
The dot is controlled by an electrostatic gate, 
and its energy levels are understood to be dominated by confinement rather than Coulomb charging energy.   
For our purposes, we thus choose to describe it as a single level dot coupled to both a chiral edge channel of 
the integer quantum Hall states, and a bosonic bath representing the electromagnetic environment, 
as illustrated in Fig.~\ref{fig:model}. The Hamiltonian is thus given by:
\begin{align}
H &= \eps_d n_d + \int \! dk \, k a_k^{\dagger} a_k + 
\sqrt{\Gamma} (d^{\dagger} \tilde{a}_0 + \tilde{a}_0^{\dagger} d) \nonumber \\
&+ \int \! dk \, k b_k^{\dagger} b_k 
+ \sqrt{2 \gamma_p} (n_d \tilde{b}_0 + \tilde{b}_0^{\dagger} n_d)~,
\end{align}
where $d$, $a_k$ and $b_k$ are annihilation operators of the dot electron, the edge-state electrons,
and the bosonic degrees of freedom, respectively, the dot occupation is $n_d = d^{\dagger} d$,
and spin degrees of freedom have been neglected assuming that electrons are fully polarized.
Here, $\varepsilon_{d}$, $\Gamma$ and $\gamma_p$ are the energy level of the quantum dot,
the decay rate of the dot electron, and the pure dephasing rate, respectively. 
\red{The velocity of the edge-channel electrons and the environment bosons has been set 
to be unity.}
We have introduced the real-space representations: 
\begin{align}
\tilde{a}_r = \frac{1}{\sqrt{2\pi}} \int \! dk \, a_k e^{ikr}~, \label{eq:2}
\\
\tilde{b}_r = \frac{1}{\sqrt{2\pi}} \int \! dk \, b_k e^{ikr}~. \label{eq:3}
\end{align} 
The integral over the 
bosonic bath is assumed to be taken in the range $-\infty < k < \infty$. This assumption
leads to simple Markov dynamics for the electron dynamics in the quantum dot:\cite{Iyoda13}
the fluctuating potential energy acting on the dot is then characterized by a white noise spectrum.
Throughout this paper, \red{we set $\hbar$ to unity}.

\section{Input-Output relations}

% Setup + Method overview
In this paper, we calculate the density matrix of single electrons (holes) injected from a quantum dot
subjected to an alternating gate voltage [shown in Fig.~\ref{fig:model}~(b)],
at zero temperature.\cite{footnote1} 
We assume that the period $T$
of the alternating field is much larger than the decay time $\Gamma^{-1}$ of the electrons
escaping the dot, and simultaneously
that the change of the field is sufficiently fast compared with $\Gamma^{-1}$.
Then, the initial state at $t=0$ in Fig.~\ref{fig:model}~(b)
is given by $|\psi(0)\rangle = |n_d=1\rangle \otimes |{\rm FS}\rangle \red{\otimes |{\rm vac}\rangle}$,
where $|{\rm FS}\rangle$ denotes the ground state of the chiral edge channel (the Fermi sea),
\red{and $|{\rm vac}\rangle$ denotes the vacuum state of the environment Hamiltonian at $t<0$, 
for which the occupation and electron hopping are fixed as $n_d=1$ and $\Gamma = 0$, respectively
(for details, see Appendix \ref{sec:appA}).}
In the following, we consider electron injection occurring in the interval $0<t<T/2$,
as the hole injection occurring in $T/2<t<T$ can be regarded as an independent dynamics that
gives the same contribution to the excess noise.
We first derive the input-output relations, and then utilize them to calculate
the various quantities which characterize the injected electrons.

% Input-output relation
For the derivation of the input-output relations, we use the equations of motion in the Heisenberg picture: 
\begin{align}
i \dot{a}_k(t) = [a_k, H] = ka_k + \sqrt{\Gamma/2\pi} d~, \\
i \dot{b}_k(t) = [b_k, H] = kb_k + \sqrt{\gamma_p/\pi} n_d~.
\end{align}
They are formally solved as:
\begin{align}
a_k(t) &= a_k(0) e^{-ikt} - i \sqrt{\frac{\Gamma}{2\pi}} \int_0^t \! dt' d(t') e^{ik(t'-t)}~, 
\label{eq:inputoutput1} \\
b_k(t) &= b_k(0) e^{-ikt} - i \sqrt{\frac{\gamma_p}{\pi}} \int_0^t \! dt' n_d(t') e^{ik(t'-t)}~,
\label{eq:inputoutput2} 
\end{align}
Transforming them into real-space representation by Eqs.~(\ref{eq:2}) and (\ref{eq:3}),
we obtain the input-output relations:
\begin{align}
\tilde{a}_r(t) &= \tilde{a}_{r-t}(0) - i\sqrt{\Gamma} \theta(r) \theta(t-r) d(t-r)~, \\
\tilde{b}_r(t) &= \tilde{b}_{r-t}(0) - i\sqrt{2 \gamma_p} \theta(r) \theta(t-r) n_d(t-r)~,
\end{align}
where $\theta(t)$ is the Heaviside step function.
For $0 < r < t$, these input-output relations combine the output field
$\tilde{a}_r(t)$ ($\tilde{b}_r(t)$) 
with the input field $\tilde{a}_{r-t}(0)$ ($\tilde{b}_{r-t}(0)$)
at the initial time. 

%-------------------------------------------------------------------------------
\begin{figure}[tbp]
\begin{center}
\includegraphics[width=0.85 \columnwidth]{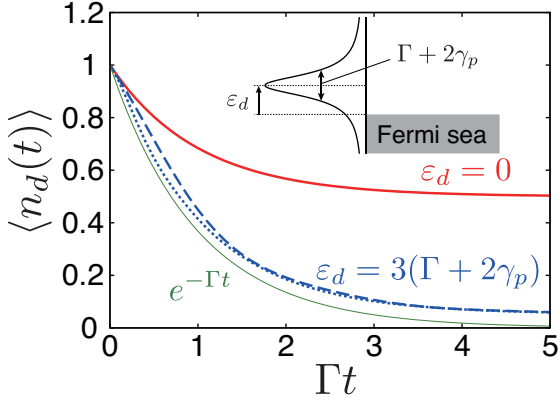}
\caption{
Time evolution of the quantum dot population $\langle n_d(t)\rangle$,
varying the dot energy $\eps_d$:
$\eps_d=0$ (red solid line),
$\eps_d=3\Gamma$ without dephasing (blue dashed line) and 
$\eps_d=3(\Gamma+2\gamma_p)$ with dephasing (blue dotted line),
where $\gamma_p = \Gamma$ is assumed. We note that for $\eps_d=0$
the population is independent of $\gamma_p$.
The exponential decay, $e^{-\Gamma t}$, is also plotted for reference (thin green solid line).
Inset: The energy diagram of the quantum dot and the Fermi sea.
The dot energy is centered at $\eps_d$ and has a linewidth of $\Gamma+2\gamma_p$.}

\label{fig:nd}
\end{center}
\end{figure}
%-------------------------------------------------------------------------------

Next, we calculate the population of the quantum dot
$\langle n_d(t) \rangle = \langle \psi(0) | n_d(t) | \psi(0) \rangle$ as a function of time.
The equation of motion for the dot operator 
is derived for $t>0$:
\begin{align}
i\dot{d}(t) = \eps_d d(t) + 
\sqrt{\Gamma} \tilde{a}_0(t) + \sqrt{2\gamma_p} 
(d(t) \tilde{b}_0(t) + 
\tilde{b}^{\dagger}_0(t) d(t))~.
\label{eq:EOMd}
\end{align}
The formal solution of this equation is obtained as:
\begin{align}
&d(t) = e^{-i\tilde{\eps}_d t} d(0)
- i\sqrt{\Gamma} \int_0^{t} \! dt' e^{i \tilde{\eps}_d(t'-t)}\tilde{a}_{-t'}(0)  \nonumber \\
&- i\sqrt{2\gamma_p} \int_0^{t} \! dt' e^{i\tilde{\eps}_d(t'-t)} (d(t') \tilde{b}_{-t'}(0) + \tilde{b}^{\dagger}_{-t'}(0) d(t'))~,
\end{align}
where $\tilde{\eps}_d = \eps_d - i\Gamma/2-i\gamma_p$.
By combining this with the input-output relations, the population of the quantum dot
is calculated as: 
\begin{align}
&\langle n_d(t) \rangle = e^{-\Gamma t} + \delta n_d(t)~, 
\label{eq:occupancy} \\
&\delta n_d(t) = \frac{\Gamma}{2\pi} \int_{-\infty}^{0} dk \frac{1}{(k-\eps_d)^2 + (\Gamma/2 + \gamma_p)^2} \nonumber \\
&\hspace{10mm} \times \left[ \frac{\Gamma + 2\gamma_p}{\Gamma} (1-e^{-\Gamma t}) + f(t) + f(t)^* \right]~, 
\label{eq:nds} \\
&f(t) = \frac{k-\eps_d+i \Gamma/2 + i \gamma_p}{k-\eps_d + i\Gamma/2 - i\gamma_p}
(e^{-\Gamma t} - e^{(-ik+i\eps_d-\Gamma/2-\gamma_p)})~.
\label{eq:nde}
\end{align}
Here, we have used the fact that $\langle \tilde{a}_{r-t}(0) \rangle = \langle \tilde{b}_{r-t}(0) \rangle = 0$, 
$\langle a^{\dagger}_k(0) a_{k'}(0) \rangle = \theta(-k) \delta(k-k')$ and
\red{$\langle b^{\dagger}_{-t'}(0) b_{-t''}(0) \rangle = 0$ ($t', t'' > 0$).}
%~\cite{footnoteadd}.}

In Fig.~\ref{fig:nd}, we plot $\langle n_d(t) \rangle$ as a function of $t$
for several values of the dot energy $\eps_d$. The population decays exponentially 
and approaches $1/2-\tan^{-1}(\eps_d/(\Gamma/2+\gamma_p))/\pi$ in the limit of $t \rightarrow \infty$.
Note that the dot population does not decay completely
due to the finite overlap in energy between the dot level and the Fermi sea.
(see the inset of Fig.~\ref{fig:nd}).
For complete injection of one electron, the condition $\eps_d \gg \Gamma/2 + \gamma_p$ is 
required.

\section{Density matrix}

The injected single electron wave packet is characterized by the density matrix:
\begin{align}
\rho(r,r',t) = \langle \tilde{a}^{\dagger}_{r'}(t) \tilde{a}_r(t) \rangle~.
\end{align}
Using the input-output relation of Eq.~(\ref{eq:inputoutput1}),
this density matrix can be separated into two parts: 
$\rho(r,r',t) = \rho_0(r,r',t) + \delta \rho(r,r',t)$, each of which is defined as:
\begin{align}
\rho_0(r,r',t) &= \langle \tilde{a}_{r'-t}^{\dagger}(0) \tilde{a}_{r-t}(0) \rangle~, \\
\delta \rho(r,r',t) &= \Gamma \theta(r) \theta(r') \theta(t-r) \theta(t-r') C(t-r,t-r')~, \label{eq:matrix1}
\end{align}
where $C(t,t') = \langle d^{\dagger}(t') d(t) \rangle$.

Our objective is to probe
the indistinguishability of electron wave packets. 
Experimentally, this is detected
by colliding two such electrons at the location of a QPC after their propagation along opposite edges
of a quantum Hall bar (as in the experiment of Ref.~\onlinecite{Bocquillon13})
and by measuring the zero frequency current-current correlations. 
This constitutes the electronic counterpart of the HOM experiment of quantum optics. 
As the first part $\rho_0(r,r',t)$ corresponds to the density matrix without electron injection,
which does not contribute to the excess noise measurement in the HOM experiment, 
we focus on the second part $\delta \rho(r,r',t)$ in the range of $0 < r < t$
and $0 < r' < t$. 
From Eq.~(\ref{eq:matrix1}), this is achieved by
calculating
the correlation function of the dot
$\langle d^{\dagger}(t') d(t) \rangle$. 
For $t'<t$, the equation of motion for this correlation function is derived:
\begin{align}
i \frac{d}{dt} C(t,t') = \tilde{\eps}_d C(t,t')
+ \frac{\Gamma}{2\pi} \int_{-\infty}^0 dk e^{-ikt}\frac{e^{ikt'}-e^{i \tilde{\eps}_d^* t'}}{k-\tilde{\eps}_d^*}~,
\end{align}
With the solution:
\begin{align}
&C(t,t') = e^{-i\tilde{\eps}_d(t-t')} \langle n_d(t') \rangle + \delta C^{>}(t,t')~,  \\ 
&\delta C^{>}(t,t') = \frac{\Gamma}{2\pi} \int_{-\infty}^0 dk 
\frac{e^{-ikt'}}{(k-\eps_d)^2 + (\Gamma/2 + \gamma_p)^2} \nonumber \\
&\hspace{12mm} \times (e^{-ik(t-t')}-e^{-i\tilde{\eps}_d(t-t')})(e^{ikt'} - e^{i\tilde{\eps}_d^* t'})~.
\end{align}
In a similar way, the correlation function is calculated for $t'>t$:
\begin{align}
&C(t,t') = e^{i\tilde{\eps}_d^* (t'-t)} \langle n_d(t) \rangle + \delta C^{<}(t,t')~,  \\ 
&\delta C^{<}(t,t') = (\delta C^{>}(t',t) )^*~.
\end{align}
By combining these results with Eqs.~(\ref{eq:occupancy})-(\ref{eq:nde}) and (\ref{eq:matrix1}),
one can calculate the density matrix $\delta \rho(r,r')$ for arbitrary sets of the parameters.

In the following discussion, we assume for simplicity that the energy level $\eps_d$ measured from the Fermi energy
is much larger than its linewidth ($\Gamma$ and $\gamma_p$) to realize the complete injection of a single 
electron. Then, $\delta n_d(t)$, $\delta C^{<}(t,t')$, and $\delta C^{>}(t,t')$, which describe
the effect of the Fermi sea in the edge channel, are evaluated:
\begin{align}
& |\delta n_d(t)| < \frac{5\Gamma/2 + \gamma_p}{\pi \eps_d}~, \label{eq:deltand} \\
& |\delta C^{<,>}(t,t')| \le \frac{2\Gamma}{\pi \eps_d}~, \label{eq:deltaC}
\end{align}
for $t, t' \ge 0$ (for derivation, see Appendix~\ref{sec:appB}), and
can be neglected when $\eps_d/\Gamma, \eps_d/\gamma_p \gg 1$, as we assume here.
The correlation function $C(t,t')$ is then calculated using $\langle n_d(t) \rangle = e^{-\Gamma t}$:
\begin{align}
C(t,t') = \left\{ \begin{array}{ll}
e^{-i \tilde{\eps}_d (t-t')-\Gamma t'}, & (t>t')~, \\
e^{i \tilde{\eps}_d^* (t'-t)-\Gamma t}, & (t<t')~.
\end{array} \right.
\end{align}
Now we switch to a frame moving at the Fermi velocity, $R=r-t$ and $R'=r'-t$. Then,
in the limit of $t \rightarrow \infty$, the density matrix becomes independent of time. It is given by:
\begin{align}
& \delta \rho(r,r',t) = \delta\rho (R', R) \nonumber \\
&= \left\{ \begin{array}{ll}
\Gamma e^{(-i\eps_d-\Gamma/2-\gamma_p)(R'-R)} e^{\Gamma R'}, & (R<R'<0)~, \\
\Gamma e^{(i\eps_d-\Gamma/2-\gamma_p)(R-R')} e^{\Gamma R}, & (R'<R<0)~, \\
0~, & ({\rm otherwise})~.
\end{array} \right.
\label{eq:DensityMatrix}
\end{align}

The density matrix includes the whole information on the injected electrons. For instance,
the shape of electron wave packet is obtained by the diagonal part of the density matrix:
\begin{align}
f(R) = \delta \rho(R,R) = \Gamma e^{\Gamma R} \theta(-R)~.
\end{align}
The injected electron has an exponential wave packet shape as expected.
Note that this shape is determined only by $\Gamma$ and is
insensitive to the dephasing rate $\gamma_p$.\cite{footnote3} 
This implies that the dephasing effects are unobservable by a simple current
measurement as achieved in Ref.~\onlinecite{Feve07}. 
Another quantity which characterizes the injected electrons
is their energy spectrum defined by:
\begin{align}
S(k,t) &\equiv \langle a_k^{\dagger}(t) a_k(t) \rangle \nonumber \\
&= \int \frac{dr dr'}{2\pi} \langle \tilde{a}^{\dagger}_{r'}(t) a_r(t) \rangle e^{-ik(r-r')}~.
\end{align}
In the limit of $t \rightarrow \infty$, the spectrum is calculated using $R$ and $R'$
as $S(k,\infty) = \int dR \ dR' \delta \rho(R,R') e^{-ik(R-R')}$. 
From the density matrix of Eq.~(\ref{eq:DensityMatrix}),
we obtain the Lorentzian lineshape:
\begin{align}
S(k) = \frac{1}{\pi} \frac{\Gamma/2 + \gamma_p}{(k-\eps_d)^2 + (\Gamma/2 + \gamma_p)^2}~.
\end{align}
Here, we note as expected that the dephasing effects on the quantum dot affect the energy broadening of 
injected single electrons, in contrast with the wave-packet shape.

%-------------------------------------------------------------------------------
\begin{figure}[tbp]
\begin{center}
\includegraphics[width=0.75 \columnwidth]{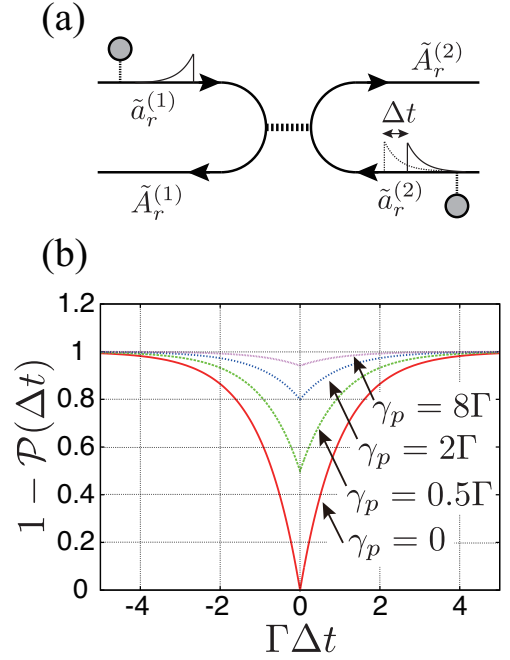}
\caption{(a) A collider setup for the fermionic Hong-Ou-Mandel experiment
using two single-electrons generators, and (b) a plot of 
$1-{\cal P}(\Delta t)$, which is proportional to the excess current noise. 
}
\label{fig:noise}
\end{center}
\end{figure}
%-------------------------------------------------------------------------------

\section{Detection of dephasing with an Hong Ou Mandel setup}

We next consider the collider setup of the HOM experiment with two 
single-electron generators as shown in Fig.~\ref{fig:noise}~(a).
We assume that the quantum point contact at the center of Fig.~\ref{fig:noise}~(a)
has a transmission (reflection) probability of $1/2$, 
playing the analog of a beam splitter in optics.
We denote the two input edge channels with $a^{(1)}_r$ and $a^{(2)}_r$, respectively.
Then, the two output edge channels are 
written as:\cite{footnote2} 
\begin{align}
A^{(1)}_r = (a^{(1)}_r + a^{(2)}_r)/\sqrt{2}~, \label{eq:outputport1} \\
A^{(2)}_r = (a^{(1)}_r - a^{(2)}_r)/\sqrt{2}~. \label{eq:outputport2}
\end{align}
In actual experiments, the measured quantity is the zero frequency excess noise:
\begin{align}
S_{ii} =\int dt \int dt' \langle \Delta I_i(t)\Delta I_i(t') \rangle~,
\end{align}
where $\Delta I_i(t) = I_i(t) - \langle I_i(t) \rangle$, 
$I_i$ ($i=1,2$) denotes the current in the output port $i$,
and ``excess'' means that the contribution of 
the Fermi sea has been subtracted out.
Because of the chiral nature of the propagation and 
the resultant propagation in the channel with a constant Fermi velocity,
defining the total number of electrons observed in 
as $N_i=\int dr (A^{(i)}_r)^{\dagger} A^{(i)}_r$, it turns out that the zero frequency 
excess current noise at the output port 1 is expressed
as $S_{11} \propto \langle (\Delta N_1)^2 \rangle$ ($\Delta N_1 = N_1 - \langle N_1 \rangle$).
Using Eq.~(\ref{eq:outputport1}) and Eq.~(\ref{eq:outputport2}), we obtain:
\begin{align}
\langle (\Delta N_1)^2 \rangle = \frac{1-{\cal P}(\Delta t)}{2}~,
\end{align}
where $\Delta t$ is the time delay between the two emitted electron wave packets and 
${\cal P}(\Delta t)$ is defined by:
\begin{align}
{\cal P}(\Delta t) = \int dr dr' \delta \rho(r,r',t) \delta \rho(r',r,t-\Delta t)~.
\end{align}
From the density matrix of Eq.~(\ref{eq:DensityMatrix}), we obtain:
\begin{align}
{\cal P}(\Delta t) = \frac{\Gamma}{\Gamma+2\gamma_p} e^{-\Gamma |\Delta t|}~.
\end{align}
In Fig.~\ref{fig:noise}~(b), we show a plot of $1-{\cal P}(\Delta t)$, which is
proportional to the excess current noise at the output port 1. If there is no dephasing
($\gamma_p=0$), the excess noise is completely suppressed for a the simultaneous collision
(time delay $\Delta t = 0$) between injected electrons. This is the manifestation of the antibunching due
to the Fermi statistics of the injected electrons. As the pure dephasing rate $\gamma_p$ increases,
the degree of antibunching is reduced, and vanishes for $\gamma_p \gg \Gamma$.
We note that ${\cal P} \equiv {\cal P}(\Delta t=0)$ corresponds to 
the purity ${\cal P}={\rm Tr} \rho^2$ of injected electrons, which has the simple form in our case:
\begin{align}
{\cal P} = \frac{\Gamma}{\Gamma + 2\gamma_p}~.
\end{align}
The purity approaches 1 for $\gamma_p \ll \Gamma$, leading to a perfect suppression
of the excess noise ($S_{11} \propto 1-{\cal P} = 0$), whereas 
the purity approaches $0$ for $\gamma_p \gg \Gamma$, producing no sign of antibunching whatsoever.

%-------------------------------------------------------------------------------
\begin{figure}[tbp]
\begin{center}
\includegraphics[width=0.75 \columnwidth]{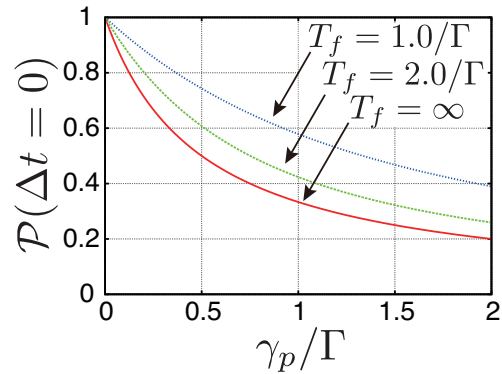}
\caption{The purity of generated single electrons after filtering
in the period of $0 < t < T_f$.}
\label{fig:filter}
\end{center}
\end{figure}
%-------------------------------------------------------------------------------

We show another useful application of the present theoretical method.
As dephasing occurs only in the quantum dot in our model, one can expect
that an electron wave-packet injected at an earlier time suffers less dephasing.
We can check this by considering wave-packets injected in the interval $0 \le t \le T_f$,
where $T_f$ is a filtering time.
This time-filtering technique is widely 
used in quantum optics, and is also implementable
in mesoscopic devices by dynamical control of the gate voltages or 
equivalently of the tunneling amplitude
between the edge channel and the dot. Then, the excess current noise is related to 
the extended purity defined by:
\begin{align}
{\cal P}_{T_f} = \frac{{\rm Tr} \rho^2}{({\rm Tr} \rho)^2} 
=\frac{\int_{-T_f}^0 dR \int_{-T_f}^{0} dR' \rho(R,R') \rho(R',R)}
{(\int_{-T_f}^0 dR \, \rho(R,R))^2}~.
\end{align}
Using our result for the density matrix, this quantity is calculated:
\begin{align}
{\cal P}_{T_f} &= \frac{2\Gamma^2}{(\Gamma + 2\gamma_p)(1-e^{-\Gamma T_f})^2} \nonumber \\
&\times \left[ \frac{1-e^{-2\Gamma T_f}}{2\Gamma} -
\frac{e^{-\Gamma T_f-2\gamma_p T_f}-e^{-2\Gamma T_f}}{\Gamma-2\gamma_p}\right].
\end{align}
In Fig.~\ref{fig:filter}, we show ${\cal P}_{T_f}$ as a function of $\gamma_p$
for $T_f = 1/\Gamma, 2/\Gamma, \infty$. Without time filtering ($T_f = \infty$),
the purity decreases monotonically with $\gamma_p$.
The purity is improved by the time filtering.
The drawback of the time filtering is 
the possibility that no electron injection occurs
from either or both of the two single-electron generators.
The probability for successfully achieving an electron collision experiment
is proportional to $({\rm Tr} \rho)^2$, 
which amounts to 0.75 (0.4) for $T_f = 2/\Gamma$ ($1/\Gamma$).
For shorter filtering times, this probability
decreases even more. This type of filtering scenario can be used to study the decoherence
source in a tunable manner in the HOM experiment.

\red{Finally, we briefly comment on other operations. In the opposite
time-filtering scheme, i.e., in considering the wave-packet injected only at $t > T_f$,
it is straightforward to show that the purity never changes from the one of
the no-filtering case. It can also be shown that if the occupied dot state is raised
at $t=0$ with $\Gamma = 0$, is kept a while up to $t=T_f$, and is relaxed 
for $t>T_f$ by turning electron hopping finite ($\Gamma > 0$), the purity 
becomes the same as the one of the no-filtering case, because such an
operation is expressed just by a shift of the origin of time.}

\section{Asymmetric wave packet collisions}

Finally, we extend the present calculations to the case of an
asymmetric wave-packet collision. In this context, asymetric can have two different
meanings: on the one hand, the injection coupling between the dots and their respective 
edge channel may differ because of limitations in nanolithography; on the other hand, 
since it is possible to adjust the dot energy levels with independent gates, controlled 
detuning is readily accessible. Both effects lead to asymetries of the shape of the electrons 
wave packets which can be analyzed with a Hong Ou Mandel interferometer. 
 
We denote the decay rate, the pure dephasing rate, and the dot energy 
by $\Gamma$, $\gamma_p$, $\eps_d$ ($\Gamma'$, $\gamma'_p$, $\eps'_d$)
for the single electron generator at the input channel 1 (2).
One then obtains the excess noise, which is proportional to $1-{\cal P}'(\Delta t)$, where
\begin{align}
{\cal P}'(\Delta t) \equiv \int dR \int dR' \rho(R,R') \rho'(R'+\Delta t,R+\Delta t)~,
\end{align}
and $\rho(R,R')$ and $\rho'(R,R')$ are the density matrix of the injected single electrons
at the input channel 1 and 2, respectively. Using previous expressions, it is easy to obtain:
\begin{align}
{\cal P}'(\Delta t) &= \left\{ \begin{array}{ll}
e^{-\Gamma' \Delta t} {\cal P}'& (\Delta t>0) \\
e^{-\Gamma |\Delta t|} {\cal P}' & (\Delta t< 0)
\end{array}
\right. \\
{\cal P}' &= \frac{2\Gamma \Gamma'}{\Gamma+\Gamma'} \times
\frac{\tilde{\Gamma}}{\tilde{\Gamma}^2 +  (\eps_d - \eps_d')^2}~,
\end{align}
where $\tilde{\Gamma} = \Gamma/2 + \Gamma'/2 + \gamma_p +\gamma_p'$.
We discuss the above result as follows. First, assuming both zero detuning and dephasing, 
but different decay rates for the two injection processes, we obtain the same result as in 
Ref.~\onlinecite{Jonckheere12}: the HOM dip is asymmetric and \red{lifted} for zero time delay. 
Second, we notice that energy detuning alone leads to the \red{lifting} of the HOM dip. 
In both cases this reflects the fact that the two wave packets are distinguishable. 
Furthermore, we see that for arbitrary parameters of the two dots, information about the decay rate of the two dots
can first be obtained by fitting the two (asymmetric) exponential sides of the dip. Knowing these decay rates, 
and by tuning the (constant) gate voltage on the dot so as to achieve $\eps_d = \eps_d'$, one could 
envision in practice to extract the quantity $\gamma_p +\gamma_p'$ which characterizes 
the dephasing to the whole setup. Alternatively, when carefully building a symmetric
setup, the measurement of the dip for zero time delay is directly related 
to the energy detuning. 

\section{Conclusion}

In summary, we have discussed how the dephasing of an electron in a quantum dot 
due to an electromagnetic environment affects the coherence of injected electrons, 
using the input-output formulation which is inspired from quantum optics. 
We showed that the density matrix of the electrons which propagate on the chiral edge
can be simply expressed in terms of the dot correlation function. 
This density matrix can be readily used to compute the fluctuations of the 
electron number -- or the zero frequency noise -- at either output of a beamsplitter
collision experiment which constitutes the electrical analog of an
HOM experiment, which is considered as a standard test of 
fundamental quantum mechanics. 
We have shown that the environmental noise does not change the spatial profile
of the emitted electrons, whereas the spectrum and the degree of antibunching is definitely affected 
by the dephasing of the quantum dot.
This treatment yields a rather simple explanation 
for the \red{lifting} of the HOM dip when the injected electrons suffer dephasing 
due to the their interaction with the electromagnetic environment located in the vicinity of 
the quantum dot. 
We have also shown that the time filtering helps
to enhance the purity, and that a generalization of the present results to an asymetric setup
is possible, with possible implication for measuring the energy detuning of the two injectors
in actual experiments.  Our approach provides not only a practical tool but also
a clear and comprehensive picture of decoherence phenomena in single electron 
generation, and the results are directly connected to past and ongoing experiments
on single electron generation which are performed in the quantum Hall effect. 

Extensions to periodic gate voltage pulses, allowing for instance to perform hole-hole or electron-hole 
collisions could be envisioned, although some modifications of the present treatment
would be required.
More complex situations of interest, such as the fractionalization of charges\cite{Wahl13,Bocquillon13b}
in quantum Hall bars containing several channels (which constitute yet another source of decoherence), or
such as the consideration of similar dephasing effects when 
the quantum dot is connected to helical edge states in topological insulators\cite{Inhofer13,Hofer13} 
constitute important developments for future research.
\red{
Note that such fractionalization scenario for decoherence as described in Ref.~\onlinecite{Wahl13} does not apply in the present work, because we are dealing here with a filling factor one of the QHE which allows the presence of a single edge state only.  In contrast, Ref.~\onlinecite{Wahl13}, which describes the experiment of Ref.~\onlinecite{Bocquillon13} at filling factor two, points out that decoherence occurs because of  Coulomb interactions with a (passive) second copropagating edge state.  While other sources of decoherence cannot be ruled out in the present geometry, we believe that environmental noise on the injecting dot is likely to be a dominant source of dephasing.}

\acknowledgements

We acknowledge helpful discussions with T. Jonckheere and J. Rech. 
T. K. was supported by JSPS KAKENHI Grant Number 24540316.
T. M. acknowledges the support of ANR-2010-BLANC-0412 (``1 shot'').
This work has been carried out in the framework of the Labex Archim\`{e}de (ANR-11-LABX-0033) and 
of the A*MIDEX project (ANR-11-IDEX-0001-02), funded by the ``Investissements d'Avenir'' 
French Government program managed by the French National Research Agency (ANR).

\appendix

\red{\section{Effect of the dot-environment coupling for $t<0$}

\label{sec:appA}
In this appendix, we show that the results of this paper are unaffected by the
dot-environment coupling before single electron injection ($t<0$).
In our paper, the electron number in the dot is fixed as $n_d = 1$ 
for $t<0$. By taking $n_d = 1$ and neglecting the electron hopping between the dot 
and a chiral edge channel, the environment for $t<0$ is described by the Hamiltonian 
\begin{align}
H &= \int dk k b_k^{\dagger} b_k + \sqrt{2\gamma_p} (\tilde{b}_0 + \tilde{b}_0^{\dagger}), \\
\tilde{b}_r &= \frac{1}{\sqrt{2\pi}} \int dk b_k e^{ikr}.
\end{align}
By introducing a displaced field operator $c_k$ defined by
\begin{align}
c_k =  b_k + \frac{1}{k}\sqrt{\frac{\gamma_p}{\pi}},
\end{align}
the Hamiltonian is diagonalized as
\begin{align}
H = \int dk k c_k^{\dagger} c_k + {\rm const.}
\end{align}
Therefore, the vacuum state $|0\rangle$ for the field operator $c_k$
is an eigenstate of $H$. The real-space representation of $c_k$ is calculated as
\begin{align}
\tilde{c}_r &\equiv \frac{1}{\sqrt{2\pi}} \int dk c_k e^{ikr} = b_r + \frac{\sqrt{2\gamma_p}}{2\pi}
\int dk \frac{e^{ikr}}{k}.
\label{eq1}
\end{align}
In order to obtain physical results, we need to replace $k$ with $k \pm i\eta$ in 
the denominator in the integral of Eq.~(\ref{eq1}),
depending on the boundary condition, where $\eta$ is a positive infinitesimal quantity.
We choose the boundary condition so that the input channel of the environment ($r<0$)
is the vacuum state of both of $\tilde{c}_r$ and $\tilde{b}_r$:
\begin{align}
\tilde{c}_r |0\rangle = \tilde{b}_r |0\rangle = 0, \quad (r<0)
\end{align}
To satisfy this condition, we choose the replacement of $k \rightarrow k-i\eta$, and obtain
\begin{align}
\tilde{c}_r = \tilde{b}_r + i \sqrt{2\gamma_p} \theta(r),
\end{align}
where $\theta(r)$ is a step function. From this result, one can see that the dot-environment
coupling affects only the output channel of the environment ($r>0$). 
Because only the input channel is relevant in our paper (we used the fact that
$b_r |0\rangle = 0$ for $r<0$), the dot-environment coupling for $t<0$ does not affect 
any results of our paper.}

\section{Derivation of Eqs.~(\ref{eq:deltand})-(\ref{eq:deltaC})}
\label{sec:appB}

In this appendix, we prove the inequalities, Eq.~(\ref{eq:deltand}) and Eq.~(\ref{eq:deltaC}).
Using $|a + b| \le |a| + |b|$, $\Gamma \ge 0$, and $\gamma_p \ge 0$, 
one can evaluate the amplitude of $\delta n_d(t)$ 
from Eqs.~(\ref{eq:nds})-(\ref{eq:nde}):
\begin{align}
|\delta n_d(t)| &\le \frac{\Gamma + 2\gamma_p}{2\pi} \int_{-\infty}^0 dk
\frac{1-e^{-\Gamma t}}{(k-\eps_d)^2 + (\Gamma/2 + \gamma_p)^2} \nonumber \\
&\hspace{5mm} + \frac{2\Gamma}{\pi} \int_{-\infty}^0 dk
\frac{|e^{-\Gamma t}-e^{-(\Gamma/2+\gamma_p)t}|}{(k-\eps_d)^2+(\Gamma/2-\gamma_p)^2} \nonumber \\
&\hspace{-2mm}< \frac{1}{\pi} \left[\frac{\pi}{2}- \tan^{-1}\left(\frac{\eps_d}{\Gamma/2+\gamma_p}\right) \right] \nonumber \\
&\hspace{-2mm} + \frac{2\Gamma}{\pi|\Gamma/2-\gamma_p|} \left[\frac{\pi}{2}- \tan^{-1}\left(\frac{\eps_d}{|\Gamma/2-\gamma_p|} \right)\right]~.
\end{align}
Using $\pi/2 - \tan^{-1}(x) \approx 1/x$ ($x \gg 1$),
Eq.~(\ref{eq:deltand}) is derived for $\eps_d \gg \Gamma, \gamma_p$, By a similar way,
one can evaluate the amplitude of $\delta C^{<}(t,t')$ and $\delta C^{>}(t,t')$:
\begin{align}
|\delta C^{<,>}(t,t')| &\le \frac{\Gamma}{2\pi} 
\int_{-\infty}^0 dk \frac{4}{(k-\eps_d)^2 + (\Gamma/2+\gamma_p)^2}~,
\nonumber \\
&\hspace{-5mm} = \frac{4\Gamma}{\pi(\Gamma + 2\gamma_p)} \left[\frac{\pi}{2} - \tan^{-1}\left(\frac{\eps_d}{\Gamma/2+\gamma_p}
\right) \right]~,
\end{align}
leading to Eq.~(\ref{eq:deltaC}) for $\eps_d \gg \Gamma, \gamma_p$.


\begin{thebibliography}{99}
\bibitem{footnote4} 
\red{The quantum dot is assumed large enough that the spacing of its levels is dominated by confinement
rather than Coulomb charging energy.}
\bibitem{footnote1} One can extend the present formulation to the finite-temperature 
case in a straightforward way.
\bibitem{footnote3} If the coupling between the quantum dot and the edge state
has an energy dependence, the dephasing may affect the shape of the wave packet.
\bibitem{footnote2} In general, there appears a phase factor in 
Eqs.~(\ref{eq:outputport1})-(\ref{eq:outputport2}). It, however,
does not affect the present results, since it is only relevant when the system 
has interference paths.
\end{thebibliography}
\end{document}